\documentclass[12pt,preprint]{aastex}

\shorttitle{VLBA Observations of the ULIRG IRAS~17208--0014}
\shortauthors{Momjian et al.}

\begin{document}

\title{VLBA CONTINUUM AND H~{\footnotesize I} ABSORPTION \\ OBSERVATIONS OF THE
\\ ULTRA-LUMINOUS INFRARED GALAXY \\ IRAS~17208--0014}

\author{Emmanuel Momjian\altaffilmark{1, 2}}
\email{momjian@pa.uky.edu}

\author{Jonathan D. Romney\altaffilmark{1}}
\email{jromney@nrao.edu}

\author{Christopher L. Carilli\altaffilmark{1}}
\email{ccarilli@nrao.edu}

\author{Thomas H. Troland\altaffilmark{2}}
\email{troland@pa.uky.edu}

\and

\author{Gregory B. Taylor\altaffilmark{1}}
\email{gtaylor@nrao.edu}

\altaffiltext{1}{National Radio Astronomy  Observatory, P O Box O, Socorro, NM
 87801.}
\altaffiltext{2}{University of Kentucky, Department of Physics and Astronomy,
 Lexington, KY 40506.}

\begin{abstract}

We present phase-referenced VLBI observations of the radio continuum emission
from, and the neutral hydrogen 21~cm absorption toward, the
Ultra-Luminous Infrared Galaxy IRAS~17208--0014. The observations were carried
out at 1362~MHz using the Very Long Baseline Array, including the phased Very
Large Array as an element. The high-resolution radio continuum images reveal a
nuclear starburst region in this galaxy, which is composed of
diffuse emission approximately $670 \times 340$~pc on the plane of the sky, and
 a number of compact sources. These
sources are most likely to be clustered supernova remnants and/or
luminous radio supernovae. Their brightness temperatures range over
(2.2--6.6)$\times 10^{5}$~K, with radio spectral luminosities between $(1-10)
\times 10^{21}~{\rm W~Hz}^{-1}$. The total VLBI flux
density of the starburst region is $\sim$52~mJy, which is about 50\% of the
total flux density detected with the VLA at arcsecond resolution. For this
galaxy, we derive a massive star formation rate of $\sim$84$\pm
13~M{_\odot}$~yr$^{-1}$, and a supernova rate of $\sim$4$\pm$1~yr$^{-1}$.
H~{\footnotesize I} absorption is detected in multiple components with optical
depths ranging between 0.3 and 2.5, and velocity widths between 58 and
232~km~s$^{-1}$. The derived column densities, assuming $T_{\rm s}=100~\rm {K}$,
range over $(10-26) \times 10^{21}$~cm$^{-2}$. The
H~{\footnotesize I} absorption shows a strong velocity gradient of
453~km~s$^{-1}$ across $0\rlap{.}''36$ (274~pc). Assuming Keplerian motion, the
enclosed dynamical mass is about {$2.3 \times 10^9~({\rm
sin}^{-2}i)~M{_\odot}$}, comparable to the enclosed dynamical mass estimated
from CO observations. \end{abstract}

\subjectheadings{galaxies: individual (IRAS~17208-0014) --- galaxies: starburst
 --- radio continuum: galaxies --- radio lines: galaxies --- supernovae:
general}

\section{INTRODUCTION}
At luminosities above $10^{11}~L{_\odot}$, infrared galaxies
become the most numerous objects in the local universe ($z \leq 0.3$) \citep
{SM96}. The trigger for the intense infrared emission appears to be the strong
interaction or merger of molecular gas-rich spirals. Galaxies at the highest
infrared luminosities ($L{_{\rm IR}}(8-1000~{\mu}m) \geq 10^{12} L{_\odot}$),
known as Ultra-Luminous Infrared Galaxies (ULIRGs), appear to be advanced merger
systems and may represent an important stage in the formation of quasi-stellar
objects \citep{SAN88}.

The bulk of the energy radiated by these sources is infrared
emission from warm dust grains heated by a central power source or sources.
The critical question concerning these galaxies is whether the dust is
heated by a nuclear starburst, or an active galactic nucleus (AGN), or a
combination of both. Mid-infrared spectroscopic studies on a sample of ULIRGs by
\citet {GEN98}, suggest that 70\%--80\% of these galaxies are powered
predominantly by recently formed massive stars, and 20\%--30\% by a
central AGN. These authors conclude further that
at least half of these ULIRGs are probably powered by
both an AGN and a starburst in a 1--2~kpc diameter circumnuclear disk or ring.
The most direct evidence to date of sub-kpc nuclear starburst regions in ULIRGs
is the discovery of luminous radio supernovae and supernova remnants in both
Arp~220 \citep {SMI98} and Mrk~273 \citep {CT00}, using VLBI observations.

In this paper, we present VLBI continuum and H~{\footnotesize I}
absorption observations on the ULIRG IRAS~17208--0014 at $z=0.0426$.
This galaxy has an infrared luminosity of $L_{\rm IR}= 2.5 \times 10^{12}
L{_\odot}$, as defined in \citet {GOL95}.

Optical images of IRAS~17208--0014 at 6550 {\AA} show two tidal tails from a
merger \citep {MM90,MUR96}, and its optical spectrum resembles that of an
H~{\footnotesize II} region \citep {VEI95,SOI00}. Its near-IR images
show a very disturbed morphology and an extended but single
nucleus, suggesting a complete merger \citep {ZL93,MUR96}.
Higher resolution near-IR images reveal numerous extremely luminous clusters in
the inner 1~kpc, with  no direct evidence of an AGN \citep {SCO00}.
Infrared observations, in general, suggest that this galaxy represents the
extreme of starburst dominated sources of this type \citep {SOI00}, and
an observational proof that a collision of galaxies can lead to a mass
distribution similar to elliptical galaxies \citep {ZL93}.

The CO (1--0) emission observations of
IRAS~17208--0014 by \citet {DS98} show a source of size $1\rlap{.}''8 \times
1\rlap{.}''6$ at full width half maximum, and a strong velocity gradient with a
change of 400~km~s$^{-1}$ over $1\rlap{.}''5$ in position angle $120^{\circ}$.
The enclosed dynamical mass estimated from the CO observation is $8.5 \times
10^{9}~M{_\odot}$.

IRAS~17208--0014 also exhibits OH megamaser activity \citep {MAR89}
at 1665, 1667, and 1720~MHz. The strongest emission is in the 1667~MHz
line, with a luminosity of $L_{\rm OH}= 10^{3}~L{_\odot}$.
MERLIN 1662~MHz radio continuum observations at $0\rlap{.}''3$ resolution
\citep {MAR89} revealed the existence of an unresolved component on the longest
baselines, with a constant visibility amplitude of $\sim$35~mJy. However, the
continuum emission from this galaxy was not detected in the 18~cm global VLBI
observations by \citet {DLLS99} to a limit of $3\sigma=210~\mu{\rm
Jy~beam}^{-1}$.

Single dish observations by \citet {MAR89} at Nan\c{c}ay showed the
existence of a very wide H~{\footnotesize I} absorption line in this galaxy,
with full velocity widths of 650 and 695~km~s$^{-1}$ at 50\% and 20\% of the
maximum depth, which is $-12$~mJy.

In this paper, we report a detailed study of the H~{\footnotesize I} 21~cm
absorption and the radio continuum emission from IRAS~17208--0014. The results
reveal the H~{\footnotesize I} gas dynamics and the distribution of compact
continuum sources, possibly composed of luminous radio supernova and/or
supernova remnants, in the nuclear region of this galaxy on sub-kpc scales. We
adopt a distance of 171~Mpc to this galaxy, assuming
${H_\circ=75}$~km~s$^{-1}$~Mpc$^{-1}$. At this distance 1~mas corresponds to
0.76~pc.

\section{OBSERVATIONS AND DATA REDUCTION}

The observations were carried out at 1362~MHz on June~4, 2001
using NRAO's \footnote{The National
Radio Astronomy Observatory is a facility of the National Science Foundation
operated under cooperative agreement by Associated Universities, Inc.}
Very Long Baseline Array (VLBA), and the
phased Very Large Array (VLA) as an element. The bandwidth of the
observations was 16 MHz, in each of right and left-hand circular polarizations,
sampled at two bits and centered at the frequency of the neutral hydrogen 21~cm
line, at a heliocentric redshift of $z=0.0426$, or $cz=12790$~km~s$^{-1}$. The
data were correlated at the VLBA correlator in Socorro, NM with 512-point
spectral resolution per baseband channel, and 2.6~seconds correlator integration
time. The total observing time was 11~hours.
An initial correlation was performed at the position reported by \citet {DLLS99}
for IRAS~17208--0014 in their 18 cm VLBI OH megamaser observations, but no
emission was detected. A second and final correlation was done at the position
reported by \citet {MAR89} in their 18 cm OH megamaser and continuum
observations with the VLA and MERLIN. This position, where strong fringes were
detected, is about 5.5 arcseconds different from the position reported by \citet
{DLLS99}.

Data reduction and analysis were performed using the Astronomical Image
Processing System (AIPS) and the Astronomical Information Processing System
(AIPS++). Table~1 summarizes the parameters of these observations.

Along with the target source IRAS~17208--0014, the compact source J1730+0024 was
observed as a phase reference with a cycle time of 120 seconds, 80 seconds on
the target source and 40 seconds on the phase reference. The source J1743--0350
was used for amplitude and bandpass calibration.

After applying {\it a priori} flagging and manually excising integrations
affected by interference, we performed amplitude calibration using
the measurements of the antenna gain and the system temperature of each station,
and bandpass calibration. Different spectral averagings generated
two data sets, a one-channel continuum and a 124-channel line data set. The
phase calibrator J1730+0024 in the continuum data set, was self-calibrated and
imaged in an iterative cycle. The self-calibration solutions of J1730+0024 were
applied on both the continuum and the line data of the target source IRAS
17208--0014.

The continuum data of IRAS17208--0014 were then deconvolved and imaged with
various spatial resolutions by tapering the visibility data, to reveal its
continuum structure. Continuum signals were mainly detected on the
short-to-moderate length baselines to the phased VLA (Y27). The H~{\footnotesize
I} absorption line was only detected on the shortest baselines to Y27, primarily
on the Y27--Pie~Town baseline. The continuum emission was subtracted from the
spectral-line visibility data. The spectral-line data were then analyzed at
various spatial and spectral resolutions, and imaged by applying a two
dimensional Gaussian taper falling to 30\% at 500~k$\lambda$. Pairs of channels
were averaged together to improve the sensitivity. The resulting velocity
resolution of the H~{\footnotesize I} images was 58~km~s$^{-1}$, and the spatial
resolution $0\rlap{.}''55 \times 0\rlap{.}''35$. An optical-depth $\tau(\nu)$
cube was calculated from the H~{\footnotesize I} absorption image cube and a
continuum image with a similar resolution as $\tau(\nu) = - {\rm ln} [ 1 -
I_{\rm line}(\nu)/I_{\rm continuum}]$.
The optical
depth cube was blanked in areas where the flux density of the background
continuum image is below 9.4\% of the peak value, i.e., less than
3~mJy~beam$^{-1}$.

\section{RESULTS AND ANALYSIS}

\subsection{{\it The Radio Continuum}}

Figure~1 is our moderate resolution continuum image of the central
region in IRAS~17208--0014 at 1362~MHz, with $80\times63$ mas ($60.8
\times 47.9$~pc) resolution, obtained by using the CLEAN algorithm as
implemented in the AIPS task ``IMAGR'', with a grid weighting
intermediate between natural and uniform (${\rm ROBUST}= 0$). A two dimensional
Gaussian taper falling to 30\% at 2~M$\lambda$ in both U and V directions was
applied on the visibility data. The radio continuum emission region has an
extent of $670 \times 390$~pc on the plane of the sky, with a total flux density
of 52~mJy. We also reduced the VLA data from these observations, which
were performed in the CnB configuration and yielded a resolution of
$9\rlap{.}''5 \times 4\rlap{.}''5$. The total flux density of the source at this
resolution is 100~mJy, consistent with the value measured by \citet
{CON96} at 1.425~GHz with the VLA-B array.

An important question regarding these observations is whether the extended
emission is likely diffuse, or composed of many point sources as might be
expected for a nuclear starburst.
Due to the incomplete UV coverage in VLBI observations, especially for a source
at $0^\circ$ declination, the CLEAN algorithm tends to generate spurious
point sources when deconvolving extended emission regions at high
resolution. On the other hand, it is plausible that the extended
emission is indeed composed of mostly faint point sources, in view of the
likely starburst nature of IRAS~17208--0014. To resolve this ambiguity, we
tested two other imaging algorithms: The Multi-Scale CLEAN and the Maximum
Entropy Method (MEM), by generating two artificial data sets in AIPS++ with the
same U-V sampling, weights and sky positions as our target source. Such
simulations give us confidence in the reliability of the results obtained from
the real observations. The first data set was composed of multiple point sources
and an extended source, and the second was composed of three extended sources.
While the MEM gave very poor images for the set with compact sources and
relatively poor results on the set with large sources, the Multi-Scale CLEAN
successfully imaged both data sets without exhibiting any tendency to generate
spurious point sources. Thus, we imaged the actual observations of our target
source at high resolution (Figure~2) using the Multi-Scale Clean algorithm as
implemented in the AIPS++ tool ``IMAGER'', with a grid weighting ${\rm ROBUST}=
0$.

Figure~2 has a resolution of $36\times33$~mas ($27.4 \times 25.1$~pc) and
represents the same region as in Figure~1. The visibility data were tapered with
a Gaussian function falling to 30\% at 6~M$\lambda$ in the U direction and
4.5~M$\lambda$ in the V direction. The continuum source is resolved and consists
of a diffuse component extending over $670\times 390$~pc, punctuated by a number
of compact sources.

There are about 25 compact sources in this high resolution image with flux
densities larger than 5$\sigma=350~\mu$Jy~beam$^{-1}$, but in Table~2 we list
only the seven strongest compact sources, with flux densities higher than
9$\sigma=630~\mu$Jy~beam$^{-1}$. We selected this conservative limit
because only
these sources were consistently reproduced in the total intensity (Stokes~I),
and in the individual RR and LL images, with comparable flux densities. The
remaining sources lacked this consistency.

The compact source parameters in Table~2 were obtained by fitting
Gaussian functions using the tool ``IMAGEFITTER'' in AIPS++. The
positions (Column~2) are relative to the peak surface brightness.
Column~3 lists the surface brightnesses of these source, and Column~4 their
integrated fluxes. Columns~5, 6 are the major and minor axes of the Gaussians at
full width half maximum, and Column~7 are their position angles.

Phase-referencing observations allow the determination of the
absolute position of the target source and its components, if any, from the
position of the calibrator \citep {WAL99}. The peak emission in
IRAS~17208--0014, which is 1043~$\mu$Jy~beam$^{-1}$, is found to be at
$\alpha\rm{(J2000)}=17^{\rm h} 23^{\rm m} 21\rlap{.}^{\rm s} 9554$,
$\delta\rm{(J2000)}=-00^{\circ} 17' 00\rlap{.}'' 938$. In applying the
phase-referencing technique, the accuracy of the calibrator position is
important. The position of J1730+0024 was obtained from the Jodrell Bank--VLA
Astrometric Survey (JVAS) of phase calibrators, with a position uncertainty of
$\sim$14~mas \citep{BRO98}. We verified the reported position of this
calibrator, as well as its structure and its VLBI flux density, with a short
VLBA observation preceding our main observations.

\subsection{{\it The} H~{\footnotesize I} {\it Absorption}}

The H~{\footnotesize I} absorption detected against the whole structure of
the continuum has a full width at 20\% maximum of 696~km~s$^{-1}$, and is
composed of several components.
Figure~3 shows Hanning smoothed spectra of H~{\footnotesize I} optical depth
averaged over various regions against the low resolution continuum image of
IRAS~17208--0014. Figure 4 presents optical depth images covering the velocity
range 13170.3--12539.0~km~s$^{-1}$. These images explicitly show the variation
of the H~{\footnotesize I} opacity against the nuclear region of
IRAS~17208--0014. In the lower panel of Figure~5, Figure~5{\it a} is the
H~{\footnotesize I} position-velocity (P-V) plot along the major axis of the
radio source (position angle $52^{\circ}$), and Figure~5{\it b} is a P-V diagram
along another cut parallel to the major axis, as shown in the continuum image at
the upper panel of Figure~5. We distinguish five main absorption peaks in both
the P-V and the optical-depth images. Their velocity widths range between 58 and
232~km~s$^{-1}$ at half maximum, with optical depths between 0.3 and 2.5.
Table~3 summarizes the physical characteristics of these five H~{\footnotesize
I} absorption features. The velocities (Column~1) refer to peaks of these
features as seen in the optical-depth spectra. The widths of these lines
(Column~2) are the approximate full widths at half peak optical depth.
$N_{\rm HI}/T_{\rm s}$ of each peak (Column~4) is calculated as $N_{\rm
HI}/T_{\rm s}~({\rm cm}^{-2}~{\rm K}^{-1}) = 1.823 \times 10^{18} \int \tau(v)
dv$. The column densities $N_{\rm HI}$ (Column 5) are derived assuming $T_{\rm
s}=100$~K. The visual extinction of each feature as a function of the
H~{\footnotesize I} spin temperature  $A_{\rm v}/T_{\rm s}$ (Column~6) is
computed assuming a Galactic dust-to-gas ratio \citep{SSD87}, and the visual
extinctions $A_{\rm v}$ (Column 7) are also derived for $T_{\rm s}=100$~K.

Our analysis show that the strongest velocity gradient in the H~{\footnotesize
I} absorption is along a position angle of $120^{\circ}$, and not along the
major axis of the radio source. The upper
panel of Figure~6 is the H~{\footnotesize I} absorption velocity field, the
contours are in steps of 50~km~s$^{-1}$ and cover a velocity range between 
12600 and 13000~km~s$^{-1}$. The lower panel of Figure~6 is a P-V plot along a
cut where the strongest velocity gradient is seen (position angle
$120^{\circ}$). This gradient is 453~km~s$^{-1}$ over 360~mas (274~pc),
i.e.~1653~km~s$^{-1}$~kpc$^{-1}$.

The spatial resolution of the low resolution
continuum image and the H~{\footnotesize I} images (Figures 3, 4, 5, \& 6) is
$0\rlap{.}''55 \times 0\rlap{.}''35$ ($418 \times 266$~pc).
The rms noise level of the H~{\footnotesize I} image cube is
0.5~mJy~beam$^{-1}$. All  these images were reconstructed with
natural grid weighting (${\rm ROBUST}= 5$ in AIPS task ``IMAGR'').

\section{DISCUSSION}

\subsection{{\it The Radio Continuum}}
At a moderate resolution (Figure~1), the nuclear region of
IRAS~17208--0014 is composed of large scale structures a few hundred
mas in size, with an average brightness temperature of 3$\times 10^{5}$~K.
Our high resolution continuum results (Figure~2) show that the nuclear region
is composed of a diffuse emission, covering an overall area
of $0\rlap{.}''88 \times 0\rlap{.}''51$ ($670 \times 340$~pc),
punctuated by multiple bright compact sources with flux densities greater than
five times the rms noise level, or $350~\mu$Jy. At the full resolution
of our array, this region is over-resolved and does not reveal
any single dominant source with very high brightness temperature. This
morphology resembles those of the starburst nuclei in M~82 \citep{MUX94,PED99},
NGC~253 \citep{UA97}, Arp~220 \citep{SMI98}, Mrk~273 \citep{CT00}, and III~Zw~35
\citep{PIH01}. The average size of these compact sources is $70 \times 48$~mas
($53 \times 37$~pc), and their brightness temperatures are between
(2.2--6.6)$\times 10^{5}$~K, indicating the emission is non-thermal and not from
H~{\footnotesize II} regions. The non-thermal emission in nuclear starburst
galaxies is usually attributed to synchrotron radiation from electrons
accelerated in supernova remnant shocks \citep{CON92}.

\subsubsection{The starburst and the star formation rate}
A remarkably tight and linear correlation between
the total radio continuum emission and the far-IR luminosities ($L_{\rm FIR}$)
is well known in ``normal'' galaxies where the main energy source is not due to
a supermassive black hole \citep{CON92}. The most obvious interpretation of this
correlation is the presence of massive stars that both provide relativistic
particles via supernova events, and heat the interstellar dust which radiates
in FIR wavelengths \citep{HSR85,WK88,CON92}.

The ratio of infrared to radio luminosity is expressed by the quantity $q$ as
defined by Helou et al.~(1985):
\begin{equation}
q={\rm log}~\{[ FIR/(3.75 \times 10^{12}]/S_{1.4~{\rm GHz}}\},
\end{equation}
where $FIR$ is given by:
\begin{equation}
FIR=1.26 \times 10^{-14}(2.58S_{60~{\mu{\rm m}}}+S_{100~{\mu{\rm m}}}).
\end{equation}
For IRAS~17208--0014, the total flux density at 1.4~GHz is 100~mJy (this work),
and the IR flux densities at 60 and 100~$\mu{\rm m}$ are $S_{60~{\mu{\rm
m}}}=34.67$~Jy and $S_{100~{\mu{\rm m}}}=37.65$~Jy, respectively \citep{SAN95}.
The resulting $q$ is 2.63, reasonably consistent with the mean value of 2.34 for
IR selected galaxies, and dispersion of $\pm 0.33$ for galaxies with $L_{\rm
60~\mu m} > 10^{11} L{_\odot}$, obtained by \citet {YRC01}. Values
of $q$ less than 1.64 would indicate the presence of a radio loud AGN \citep
{YRC01}. Thus, based on the radio-FIR correlation, it would appear that
IRAS~17208--0014 is dominated by a starburst.

An estimate of the massive star formation rate and the supernova rate for a
starburst galaxy can be obtained in two independent ways. The first is a
theoretical approach which assumes an initial mass function (IMF) \citep{SS91},
and the second is based on an empirical relationship between the observed
non-thermal radio flux density and the supernova rate in the
Galaxy \citep{CY90}.

For the theoretical approach, we assume a modified Miller-Scalo IMF of the form
$\psi(M)\propto M^{-5/2}$. This function is truncated at ${M_{\rm l}} \leq
1~M_\odot$ and ${M_{\rm u}} \geq 100~M_\odot$ \citep{SCA86,CON92,SLL98}. The
models presented by \citet{SS91} give the following relationship between the
luminosity of the starburst and the star formation rate ($SFR$):
\begin{equation} L_*=1.18 \times 10^{10} \Big(\frac{M_{\rm
l}}{1~M_\odot}\Big)^{\alpha} \Big(\frac{M_{\rm u}}{45~M_\odot}\Big)^{0.37}
\Big(\frac{SFR}{1~M_\odot~{\rm yr}^{-1}}\Big)
\Big(\frac{t_{\rm B}}{10^8~{\rm yr}^{-1}}\Big)^{0.67}L_\odot,
\end{equation}
where ${M_{\rm l}}$ and ${M_{\rm u}}$ are the lower and upper mass limits for
star formation, respectively, ${t_{\rm B}}=10^8~{\rm yr}$ is the lifetime of the
starburst, and $\alpha=0.23$ and 0.55 for ${M_{\rm l}}<1~M_\odot$ and
$>1~M_\odot$, respectively.

The supernova rate $\nu_{\rm SN}$ of a starburst galaxy can be calculated from
the star formation rate using the following expression from \citet{SMI98},
that assumes the same IMF introduced above:
\begin{equation}
\nu_{\rm
SN}\simeq0.33(SFR)\frac{(M_u^{-3/2}-M_{SN}^{-3/2})}{(M_u^{-1/2}-M_l^{-1/2})}~~~
~~{\rm yr}^{-1}, \end{equation}
where $M_{\rm SN}=8~M_\odot$ for Type II supernovae.

To obtain the star formation and supernova rates in IRAS~17208--0014
($L_{\rm FIR}=2.3\times 10^{12} L{_\odot}$),
we use the lower and upper mass arguments obtained for Arp~220, the
prototype ULIRG which has a very similar $q$ value to our galaxy. For
Arp~220 an upper limit mass was set from free-free emission and Pa$\beta$
observations and the lower mass limit was set from the CO dynamical mass
measurements and the adopted disk model (\citet{SMI98} and
references therein).
These limits are ${M_{\rm l}}=5~M_\odot$ and ${M_{\rm u}}= 28~M_\odot$.
It has also been found that the far-IR luminosity is a good measure of the
bolometric luminosity produced by fairly massive $(M \geq 5M_\odot)$
young stars \citep{CON92}. Thus, with these
mass limits, we can state that $L_*= L_{\rm FIR}$.
The resulting star formation rate (Equation~3) is
$SFR\simeq96~M_\odot~{\rm yr}^{-1}$, and the supernova rate (Equation~4) is
$\nu_{\rm SN}\simeq4.6~{\rm yr}^{-1}$.

As an alternative, we can derive these rates using the empirical approach
of \citet{CY90}, which is based on the observed Galactic non-thermal luminosity
($L_{\rm NT}$) and the supernova rate $\nu_{\rm SN}$. This relationship is
\begin{equation}
\Big(\frac{L_{\rm NT}}{10^{22}~{\rm W~Hz}^{-1}}\Big)=13
\Big(\frac{\nu}{\rm GHz}\Big)^{-\alpha}
\Big(\frac{\nu_{\rm SN}}{{\rm yr}^{-1}}\Big),
\end{equation}
where $\alpha\sim0.8$ is the non-thermal spectral index for ``normal'' galaxies.
Our 52~mJy total VLBI flux density for IRAS~17208--0014, which is
mostly non-thermal, corresponds to a luminosity of $1.81 \times 10^{23}~{\rm
W~Hz}^{-1}$. From Equation~(5), we get $\nu_{\rm SN}\simeq1.8~{\rm yr}^{-1}$,
and from Equation(4) we get a massive star formation rate of $37~M_\odot~{\rm
yr}^{-1}$, assuming the mass constraints as for Arp~220.

These values are lower than the rates obtained from the far-IR luminosity by
a factor of 2.6, and suggest that there might be non-thermal diffuse
emission which is not detected by our VLBI array. As shown by \citet{CON92}, the
observed fraction of thermal emission in ``normal'' galaxies is $\leq 0.1$ at
1.4~GHz. Thus, a reasonable approximation would be to consider the
total VLA flux density, which is 100~mJy, to be non-thermal.
Then, following the above argument with the luminosity $3.48 \times 10^{23}~{\rm
W~Hz}^{-1}$ that corresponds to the detected VLA flux density, we obtain
$\nu_{\rm SN}\simeq3.4~{\rm yr}^{-1}$ and $SFR\simeq71~M_\odot~{\rm yr}^{-1}$.
These values are more consistent with the rates obtained from the far-IR
luminosity using similar mass limits, and support the suggestion that there is
undetected diffuse flux in our VLBI observations.

We averaged the rates obtained from both the
far-IR luminosity and the VLA radio flux density to obtain our best estimates
for the massive star formation rate in IRAS~17208--0014,
$\sim$$84\pm13~M_\odot~{\rm yr}^{-1}$, and the supernova rate $\sim$$4\pm1~{\rm
yr}^{-1}$.
The derived star formation and supernova rates are consistent with
the rates in the ULIRG Arp 220, where \citet{SMI98} obtained a
star formation rate of $50-100~M_\odot~{\rm yr}^{-1}$, and a
supernova rate of $1.75-3.5~{\rm yr}^{-1}$
in VLBI observations at 18 cm.
Our supernova rate is much higher than the upper
limit of 0.3~yr$^{-1}$ that \citet{UA97} derived for the nearby starburst galaxy
NGC~253 from higher-frequency VLA observations.
Follow-up VLBI observations will make it possible to look
for new sources (i.e.~RSNe) to directly constrain both the supernova and the
massive star formation rates in the nuclear starburst region of
IRAS~17208--0014.

An estimate of the magnetic field strength and the pressure of the relativistic
electrons responsible for the synchrotron radiation detected from the
starburst region can be obtained by minimizing the summed energy in magnetic
fields and relativistic particles \citep {MIL80}.
The resulting magnetic field is $\sim$$144~\mu$Gauss, and the corresponding
pressure is $\sim$$6 \times 10^{-10}$~dyn~cm$^{-2}$, indicating extreme physical
conditions compared to the disks of spiral galaxies.

\subsubsection{The nature of the compact sources}
The VLBI continuum emission in IRAS~17208--0014 extends over a
region of $670 \times 340$~pc, and can be explained in part by synchrotron
radiation from aged supernova remnants (SNRs), but this region is also
punctuated by a number of bright compact sources (Figure~2). The radio spectral
luminosities of these sources range over $(1-10) \times 10^{21}~{\rm W~Hz}^{-1}$
at 1.362~GHz, and are at least an order of magnitude greater than the brightest
radio supernovae (RSNe) seen in M~82 \citep{MUX94,PED99}, but comparable to the
rare class of extreme luminosity RSNe characterized by SN~1986J in NGC~891
\citep{RUP87} and SN~1979C \citep{WS88}. More recently, a substantial population
of such luminous RSNe was discovered in Arp~220 by \citet{SMI98}, who
suggest that the high luminosities of those RSNe may indicate denser
environments, more massive progenitors, or stronger magnetic fields relative to
typical RSNe.

The compact sources revealed in our observations are a few times brighter
and larger in size than the RSNe reported in the ULIRGs Arp~220 \citep{SMI98}
and Mrk~273 \citep{CT00}. Thus, we presume that the compact
sources in the starburst region of IRAS~17208-0014 are clustered young SNRs
and/or luminous RSNe. Support for the clumpy nature of the nuclear region in
IRAS~17208--0014 is provided by the high resolution near-IR images obtained with
the Hubble Space Telescope NICMOS camera \citep{SCO00}. These images, with
resolutions between $0\rlap{.}''11$ and $0\rlap{.}''22$, reveal numerous
extremely luminous clusters of massive stars in the inner 1~kpc region of this
galaxy.

Smith et al.(1998a) have fitted smooth Gaussian and clumped models of starburst
distributions to the observed visibility functions of 11 luminous and
ultra-luminous IR galaxies. Their first conclusion was that starbursts with
standard RSNe are incapable of producing the radio power and structure of these
galaxies. For luminous RSNe models, the Gaussian fits represented the
visibility function poorly, but the clustered models, which assumed
simultaneous detonation of the RSNe within one clump, were more successful.
Excluding poor solutions, the sizes of these clumps ranged over 1--15~pc in
diameter, and the number of supernovae per clump was between 2 and 12. These
clumps are at least two times smaller than the minimum sizes implied by the
Gaussian fits to our compact sources (Table~2).

However, we cannot rule out the possibility that the compact sources in our
high resolution image are mainly powered by very luminous individual RSNe.
\citet{SMI98} fit the following expression for the post-maximum light curve
of luminous RSNe:
\begin{equation}
S_\nu = S_{\nu,{\rm max}}\Big(\frac{\Delta t}{3~{\rm yr}^{-1}}\Big)^{1.3},
\end{equation}
where ${\Delta t}=t-t_o$, and $t=t_o$ is the detonation time.
Following their discussion, individual RSNe will lie above our $5\sigma$
detection level for 7 years. Given our adopted supernova rate of 4~yr$^{-1}$, we
expect to see about 28 individual luminous RSNe, in remarkable agreement with
the 25 compact sources revealed in our observations.

Future multi-epoch VLBI observations would determine which of the two
above assumptions better explain the nature of the compact components in
IRAS~17208--0014.

\subsection{{\it The} H~{\footnotesize I} {\it Absorption}}
The wide H~{\footnotesize I} absorption observed against the
whole extent of the nuclear region in IRAS~17208--0014, with a full width
of $696~{\rm km~s}^{-1}$ at 20\% maximum, is common in ULIRGs. Similar wide
absorption lines have been found in Arp~220 \citep{MS88}, Mrk~273 \citep{BOT85},
and Mrk~231 \citep{DIC82}. The present observations are consistent with the
Nan\c{c}ay 300~m meridian telescope observations by \citet{MAR89}, who found a
broad H~{\footnotesize I} absorption ($\Delta V_{\rm 20\%}=695~{\rm km~s}^{-1}$)
centered at $12790\pm10~~{\rm km~s}^{-1}$. The detected peak absorption flux
density in our observations is $-8.65$~mJy, about 72\% of the single dish
detection.

Our observations reveal the details of the H~{\footnotesize I}
absorption against the radio emission in IRAS~17208--0014 (Figures~3, 4, 5, \& 
6). The absorption is complex, and composed of five main
features (Table~3). Three of them are wide
($\Delta V_{\rm FWHM}=174-232~{\rm km~s}^{-1}$) with $\tau < 1$, and the other
two relatively narrow ($\Delta V_{\rm FWHM}=58~{\rm km~s}^{-1}$) with $\tau
> 1$. These two narrow lines appear to be localized features on the
east edge and the north-east side of the continuum emission region (Figure~4),
suggesting the existence of quiescent H~{\footnotesize I} clouds in the
ISM of the galaxy. The linear extent of these two features is $0.3-0.46$~kpc,
as seen in the optical depth images (Figure~4).

The P-V diagram in position angle 120$^{\circ}$ (Figure~6-{\it lower panel}),
shows a strong velocity gradient of 453~km~s$^{-1}$ over $0\rlap{.}''36$
(274~pc). Assuming Keplerian motion, the enclosed dynamical mass is {$2.3 \times
10^9~({\rm sin}^{-2}i)~M{_\odot}$}, where $i$ is the inclination angle. The
rotational behavior of the H~{\footnotesize I} and the derived dynamical mass
are largely consistent with the results of CO $(1-0)$
emission line observations \citep{SOL97,DS98}. The interferometric
observations of the CO $(1-0)$ by \citet{DS98} with IRAM at Plateau de
Bure, show a velocity gradient of 400~km~s$^{-1}$ over $1\rlap{.}''5$ at
position angle 120$^{\circ}$, and an enclosed mass of $8.5 \times
10^9~M{_\odot}$. The resolution of these observations was $5\rlap{.}''1 \times
1\rlap{.}''6$ in position angle 47$^{\circ}$.
In both H~{\footnotesize I} absorption and CO $(1-0)$ emission, the strongest
velocity gradient is seen along the same direction (position angle
120$^{\circ}$). The H~{\footnotesize I} absorption results show a stronger
velocity gradient (1653~km~s$^{-1}$~kpc$^{-1}$) comparing to the CO $(1-0)$
(346~km~s$^{-1}$~kpc$^{-1}$), suggesting that the H~{\footnotesize I} disk
represents the inner region of a larger molecular disk.
A similar model was proposed for the absorbing neutral hydrogen in Mrk~231, 
where the observed H~{\footnotesize I} disk was identified as the inner region
of the molecular gas seen in CO $(1-0)$ emission \citep{CWU98}.

It is likely that both atomic and molecular gas exist in the region where the
H~{\footnotesize I} absorption is detected. The emission detected in low
spatial resolution (FWHM $5\rlap{.}''1 \times 1\rlap{.}''6$) CO
observations arises from larger scales, and would not show variations on the
scales seen in the H~{\footnotesize I} absorption in IRAS~17208--0014.

The neutral hydrogen column densities of the observed absorption features are
high, covering a range $(10.1-28.0)~T_{\rm s}{\rm (K)} \times 10^{19}~{\rm
cm}^{-2}$. The corresponding visual extinctions $A_{\rm v}$, assuming a
Galactic dust-to-gas ratio, are between 0.066~$T_{\rm s}{\rm (K)}$ and
0.174~$T_{\rm s}{\rm (K)}$~mag. The discussion in \citet{SCO98} for
Arp~220, with the infrared magnitudes obtained for IRAS~17208-0014
\citep{SCO00}, implies a visual extinction value of $\sim$22~mag, comparable
to the values derived from our H~{\footnotesize I} column densities with a spin
temperature of $T_{\rm s}\sim$100 K.

Our results show a very good agreement between the physical properties
of the neutral H~{\footnotesize I} in IRAS~17208--0014 and in the other ULIRGs
previously observed in detail,
namely Arp~220 \citep{MFP01}, Mrk~273 \citep{COL99,CT00} and Mrk~231
\citep{CWU98}. The strong velocity gradients
seen in the H~{\footnotesize I} absorption of these galaxies suggest rapidly
rotating disks which are obscuring the radio-emitting nuclear regions, while the
multiple absorption features seen against different regions of the continuum
indicate the existence of several discrete clouds in those disks.

\section{CONCLUSIONS}

We have presented the results of phase-referenced VLBI observations, using the
VLBA and the phased VLA, of the 21~cm continuum emission and the
H~{\footnotesize I} absorption in the central $\sim$1.1~kpc of the very advanced
merger galaxy IRAS~17208--0014.

The high resolution continuum images reveal the details of the previously
undetected nuclear starburst region of this galaxy.
Both diffuse and compact continuum emission are detected. The total VLBI flux
is 52~mJy, and represents only half of the total flux seen with the VLA at a
lower resolution, suggesting the existence of diffuse
emission not detected by our VLBI array.
The compact sources in the starburst region are more likely clustered luminous
radio supernovae and supernova remnants, considering the H~{\footnotesize II}
spectra observed in this galaxy and its morphological similarities to other
well-known starbursts. However, we cannot rule out the possibility that each of
our compact sources is mainly powered by an individual bright RSN nested in a
region that contains faded SNRs. The brightness temperatures of the compact
structures seen in these observations are $>10^5$~K.
The flux density of the brightest source in the starburst region is less than
3\% of the total radio flux density. The results suggest that there is no
radio-loud AGN in the nuclear region of IRAS~17208--0014.

Both the far-infrared luminosity and the radio continuum flux of
this galaxy imply a massive star formation rate of $\sim$$84~M_\odot~{\rm
yr}^{-1}$ and supernova rate of $\sim$$4~{\rm yr}^{-1}$.
The estimated individual luminous RSN number agrees surprisingly well with the
number of the compact sources detected above 5$\sigma$ level.
From the minimum energy condition, we have estimated the pressure to be $\sim$$6
\times 10^{-10}$~dyn~cm$^{-2}$ and the magnetic field to be $\sim$$144~\mu$Gauss
in the starburst region.

The very wide H~{\footnotesize I} absorption is composed of five
components with velocity widths between 58 and 232~km~s$^{-1}$.
The column densities of these absorption peaks are on the order of
$10^{20}~T_{\rm s}{\rm (K)}$, and the derived visual extinctions are between
$0.066~T_{\rm s}{\rm (K)}$ and $0.174~T_{\rm s}{\rm (K)}$~mag. The strongest 
velocity gradient in the H~{\footnotesize I} absorption is seen in position
angle 120$^{\circ}$, as in the CO $(1-0)$ emission. The H~{\footnotesize I}
position-velocity profile in this position angle shows a velocity gradient of
1653~km~s$^{-1}$~kpc$^{-1}$. The calculated dynamical mass, assuming Keplerian
motion, is {$2.3 \times 10^9~({\rm sin}^{-2}i)~M{_\odot}$}, comparable to the
enclosed mass obtained from CO observations. The H~{\footnotesize I} absorption
results suggest the existence of a neutral gas disk containing several clouds,
situated inside a larger scale molecular disk as seen in the interferometric CO
$(1-0)$ emission line observations.

\section{ACKNOWLEDGMENTS}
This research has made use of the NASA/IPAC Extragalactic Database (NED) which
is operated by the Jet Propulsion Laboratory, California Institute of
Technology, under contract with the National Aeronautics and Space
Administration.
E.~M. is grateful for support from NRAO through the Pre-doctoral
Research Program.
T.~H.~T. and E.~M. acknowledge NSF support through grant AST~99-88341.

\clearpage
\begin{figure}
\epsscale{0.8}
\plotone{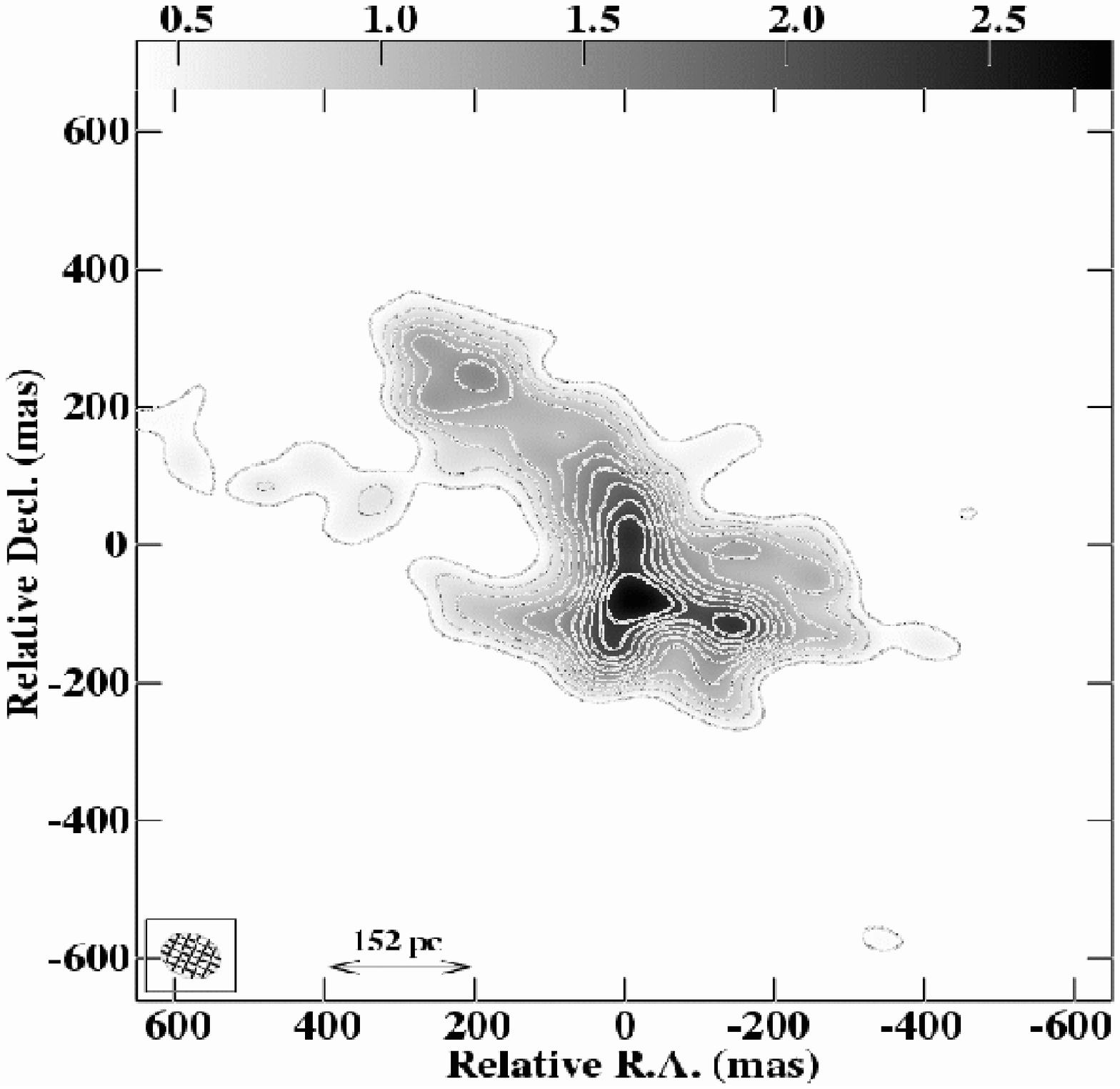}
\caption{Low resolution continuum image of the central region in
IRAS~17208--0014 at 1362~MHz. The restoring beam size is $80 \times 63$~mas in
position angle $67^{\circ}$. The peak flux is 2.8~mJy~beam$^{-1}$, and the
contour levels are at $-$4, 4, 8,$\ldots$24 times the rms noise level, which is
0.1~mJy~beam$^{-1}$. The gray scale range is indicated by the wedge at
the top of the image in units of mJy~beam$^{-1}$. The reference position
(0,0) is $\alpha \rm{(J2000)}=17^{\rm h} 23^{\rm m} 21\rlap{.}^{\rm s} 9648$,
$\delta \rm{(J2000)}=-00^{\circ} 17' 00\rlap{.}'' 819$. A two
dimensional Gaussian taper falling to 30\% at 2~M$\lambda$ was
applied.\label{FIG1}} \end{figure}

\begin{figure}
\epsscale{1}
\plotone{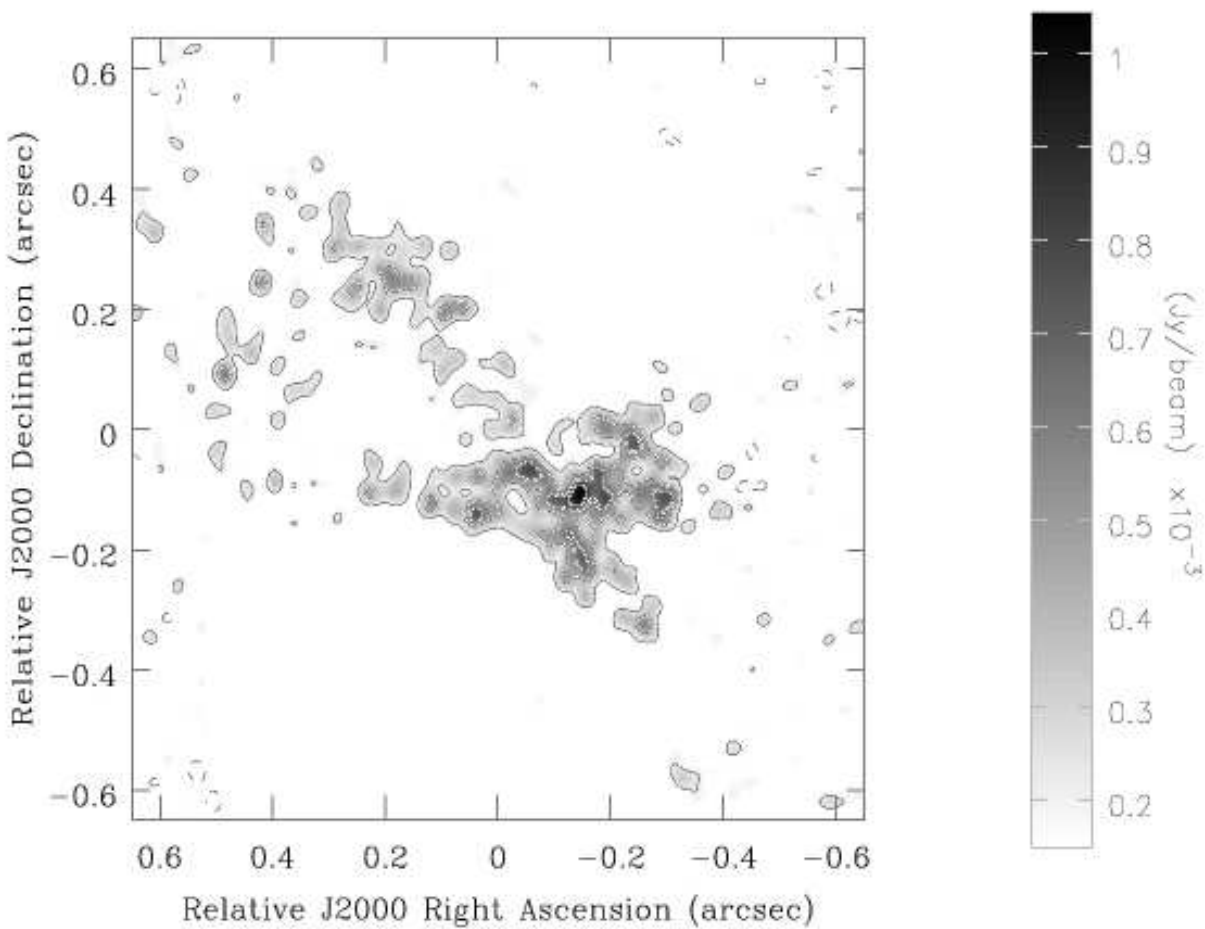}
\caption{Continuum image of the central region in IRAS~17208--0014 at 1362~MHz.
The restoring beam size is $36 \times 33$~mas in position angle
$25^{\circ}$. The peak flux is 1016~$\mu$Jy~beam$^{-1}$, and the contour levels
are at $-$3, 3, 6, 9, and 12 times the rms noise level, which is
70~$\mu$Jy~beam$^{-1}$. The gray scale range is indicated by the wedge at
the right side of the image. The reference position
(0,0) is $\alpha \rm{(J2000)}=17^{\rm h} 23^{\rm m} 21\rlap{.}^{\rm s} 9648$,
$\delta \rm{(J2000)}=-00^{\circ} 17' 00\rlap{.}'' 819$. A Gaussian taper
falling to 30\% at 6~M$\lambda$ in the U direction and 4.5~M$\lambda$ in the V
direction was applied.\label{FIG2}}
\end{figure}

\begin{figure}
\epsscale{1.0}
\plotone{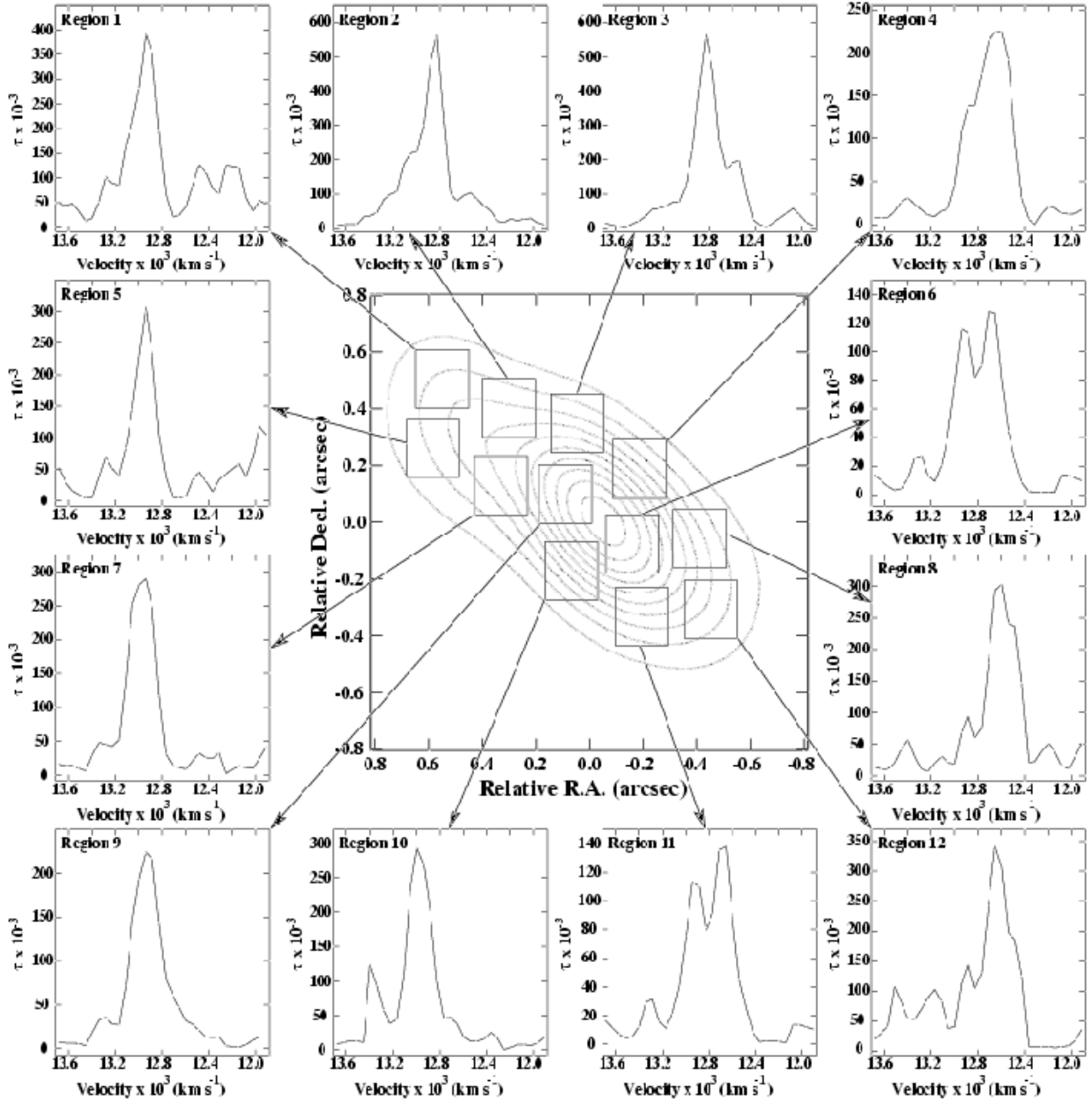}
\caption {Hanning smoothed H~{\footnotesize I} optical depth spectra obtained
at various locations against the background continuum source IRAS~17208-0014
at 1362~MHz. All the spectra are scaled to the maximum H~{\footnotesize I} 
absorption optical depth of each region.
The restoring beam size is $0\rlap{.}''55 \times 0\rlap{.}''35$ in
position angle $34^{\circ}$.
The contour levels of the
continuum image are at 3, 6,$\ldots$30~mJy~beam$^{-1}$, and the peak flux is
32~mJy~beam$^{-1}$. The reference position (0,0) is $\alpha
\rm{(J2000)}=17^{\rm h} 23^{\rm m} 21\rlap{.}^{\rm s} 9648$, $\delta
\rm{(J2000)}=-00^{\circ} 17' 00\rlap{.}'' 819$.
\label{FIG3}}
\end{figure}

\begin{figure}
\epsscale{0.75}
\plotone{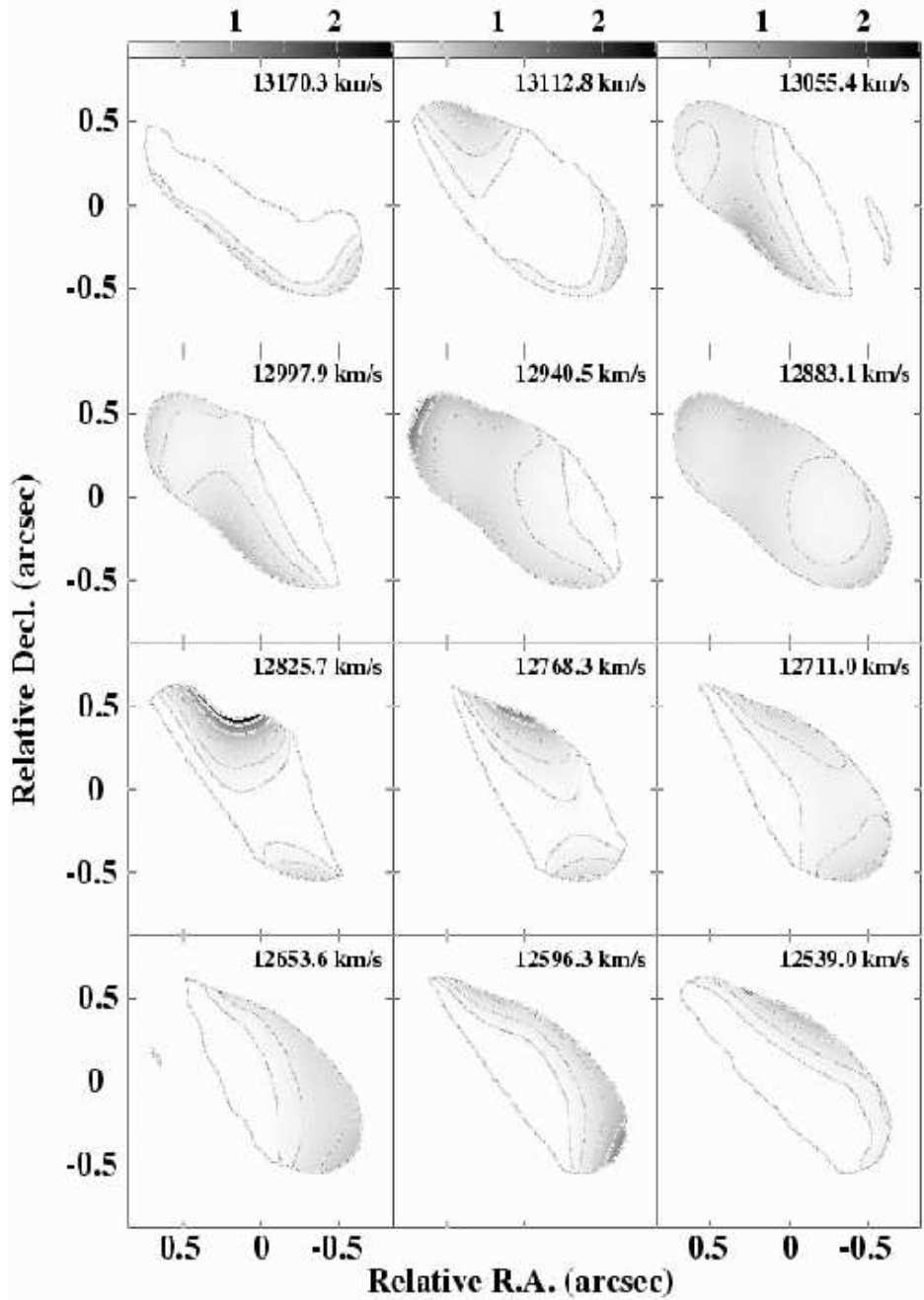}
\caption{Gray-scale and contour H~{\footnotesize I} optical depth channel
images toward IRAS~17208--0014 in the velocity range
13170.3--12539.0~km~s$^{-1}$. The restoring
beam in these images is $0\rlap{.}''55 \times 0\rlap{.}''35$ in position angle
$34^{\circ}$. The velocity resolution is 58~km~s$^{-1}$. The gray-scale range
is indicated by the step wedge at the top of the images; the contour levels are
0.1, 0.2, 0.4, 0.8, and 1.6. \label{FIG4}}
\end{figure}

\begin{figure}
\epsscale{1}
\plotone{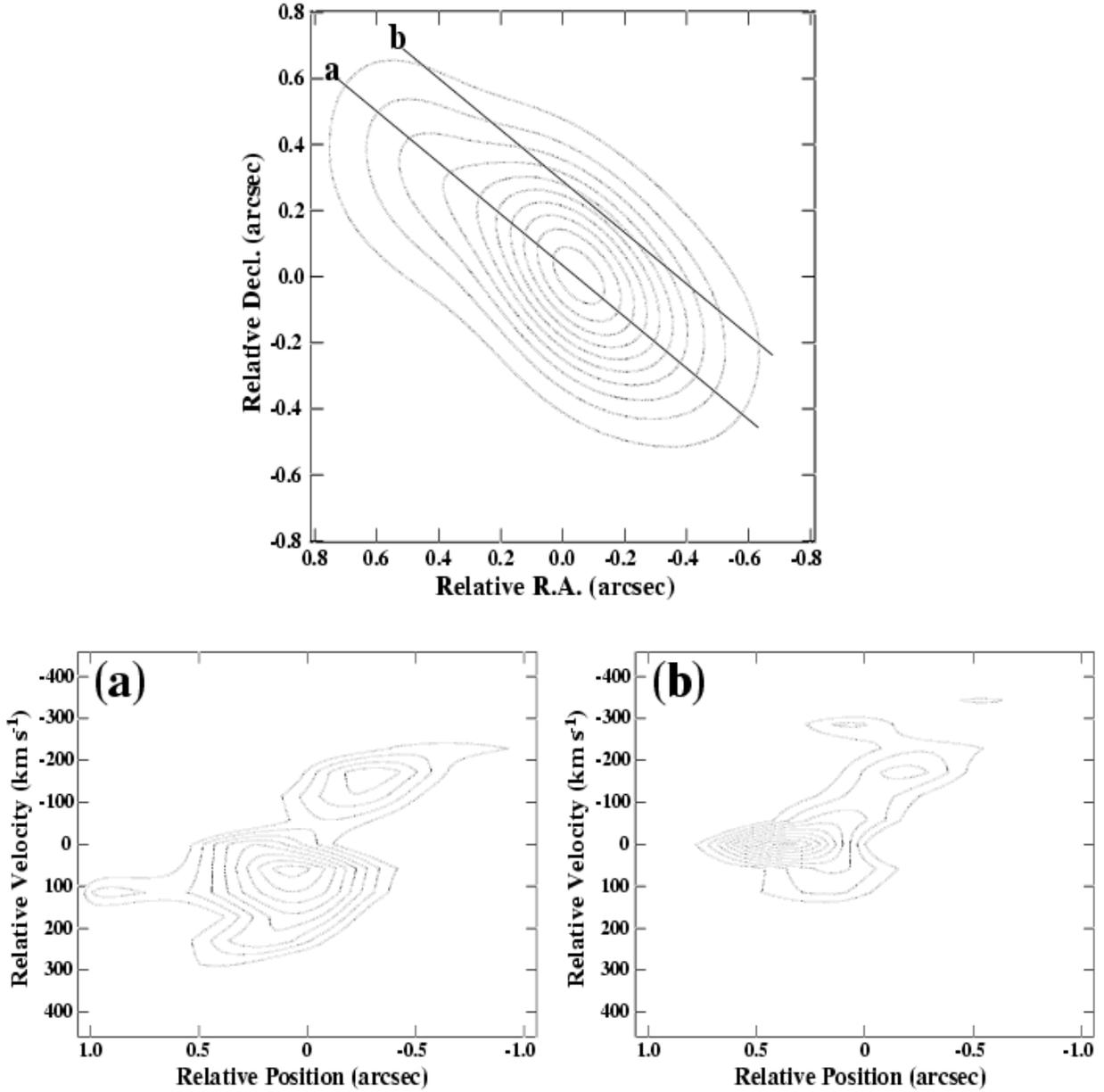}
\caption {Position-velocity plots of the H~{\footnotesize I} 21~cm absorption
along two cuts in position angle $52^{\circ}$. The contour levels are at
$-2, -2.5,\ldots-6$~mJy~beam$^{-1}$. The velocity resolution is
58~km~s$^{-1}$. The zero point on the velocity scale corresponds to a
heliocentric velocity ($cz$) of 12825.7~km~s$^{-1}$. The contour levels of the
continuum image shown at the top are at 3, 6,$\ldots$30~mJy~beam$^{-1}$, and the
peak flux is 32~mJy~beam$^{-1}$. The reference position (0,0) is $\alpha
\rm{(J2000)}=17^{\rm h} 23^{\rm m} 21\rlap{.}^{\rm s} 9648$, $\delta
\rm{(J2000)}=-00^{\circ} 17' 00\rlap{.}'' 819$.
The restoring beam size is $0\rlap{.}''55 \times 0\rlap{.}''35$ in position
angle $34^{\circ}$.
\label{FIG5}}
\end{figure}

\begin{figure}
\epsscale{0.6}
\plotone{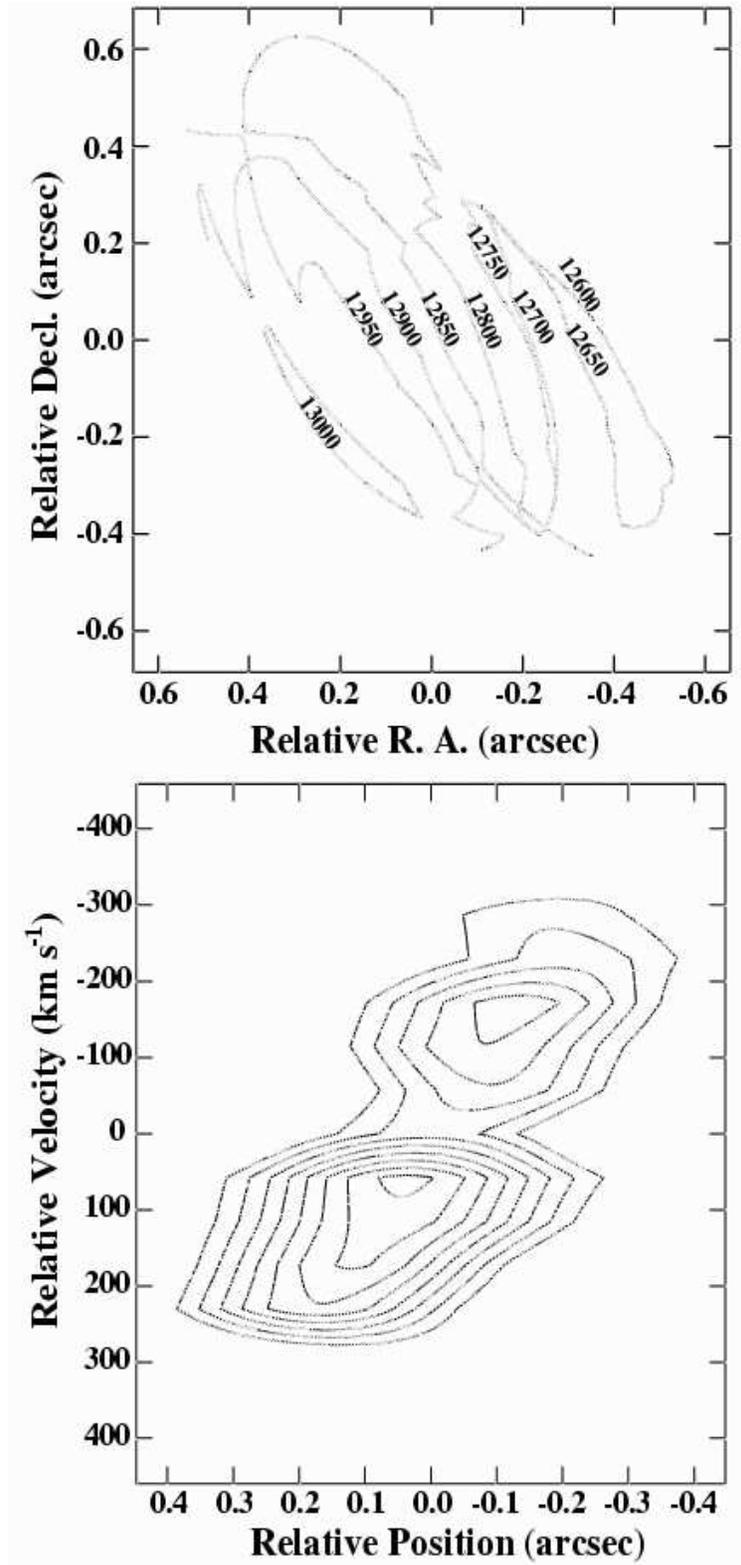}
\caption {{\it Upper panel:} Velocity contours of the H~{\footnotesize I}
21~cm absorption in steps of 50~km~s$^{-1}$. The reference position (0,0) is
$\alpha \rm{(J2000)}=17^{\rm h} 23^{\rm m} 21\rlap{.}^{\rm s} 9648$, $\delta
\rm{(J2000)}=-00^{\circ} 17' 00\rlap{.}'' 819$.
{\it Lower panel:} H~{\footnotesize I} position-velocity plot in
position angle $120^{\circ}$. The contour levels are at $-1.5,
-2,\ldots-6$~mJy~beam$^{-1}$. The
zero point on the velocity scale corresponds to a heliocentric velocity ($cz$)
of 12825.7~km~s$^{-1}$.
\label{FIG6}}
\end{figure}
\clearpage

\begin{deluxetable}{lccc}
\tablecolumns{4}
\tablewidth{0pc}
\tablecaption{P{\footnotesize ARAMETERS} {\footnotesize OF THE} VLBI
O{\footnotesize BSERVATIONS} {\footnotesize OF} IRAS~17208--0014}
\tablehead{
\colhead{Parameters} & \colhead{}  & \colhead{} &
\colhead{Values}} \startdata
Observing Date \dotfill & &  & 2001 June~4 \\
Observing Array \dotfill & &  & VLBA + Y27 \\

R.A. (J2000)\dotfill & & &  17 23 21.9554 \\
Dec. (J2000)\dotfill & & &  $-$00 17 00.938 \\

Total observing time (hr)\dotfill &  & & 11 \\

Phase-referencing cycle time (min)\dotfill &  & &  2 \\

Frequency (MHz)\dotfill & & &  1362 \\

Bandwidth (MHz)\dotfill  & & & 16 \\

Continuum image rms ($\mu$Jy beam$^{-1}$)\dotfill  & & & 70 \\

Line velocity resolution (km s$^{-1}$)\dotfill & & & 58 \\

Line image rms (mJy beam$^{-1}$)\dotfill & & & 0.5 \\

Optical depth image cutoff (mJy beam$^{-1}$)\ldots \ldots &  & &
3 \\

\enddata
\tablecomments{Units of right ascension are hours, minutes, and seconds, and
units of declination are degrees, arcminutes, and arcseconds.}
\end{deluxetable}

\clearpage

\begin{deluxetable}{cccccccc}
\rotate
\tablecolumns{8}
\tablewidth{0pc}
\tablecaption{C{\footnotesize OMPACT} S{\footnotesize OURCES} {\footnotesize IN}
IRAS~1720--0014}
\tablehead{
\colhead{}    &\colhead{}    &\colhead{}
&\multicolumn{4}{c}{Gaussian Component Parameters} \\ \cline{4-8} \\
\colhead{Source} &  & \colhead{Relative Position\tablenotemark{a}}&
\colhead{Peak\tablenotemark{b}} & \colhead{Total} &
\colhead{Major Axis\tablenotemark{c}} & \colhead{Minor Axis\tablenotemark{c}} &
\colhead{P.A.}\\
\colhead{} & \colhead{} &  \colhead{(mas)} &
\colhead{($\mu$Jy~beam$^{-1}$)} & \colhead{(mJy)} & \colhead{(pc)}&
\colhead{(pc)}& \colhead{($^{\circ}$)} \\
\colhead{(1)} & &
\colhead{(2)}& \colhead{(3)} & \colhead{(4)} & \colhead{(5)} & \colhead{(6)} &
\colhead{(7)}}
\startdata
1\dotfill & &   0 ,  0  & $1043.3 \pm 1.0$&$2.764 \pm 0.003$ & $52.3 \pm 0.4$&$
34.9 \pm 0.3$& 155\\
2\dotfill & &  87E, 40N & $791.1  \pm 0.9$&$2.356 \pm 0.003$ & $50.7 \pm 0.5$&$
40.5 \pm 0.4$& 78\\
3\dotfill & & 153W,  6S & $759.6  \pm 1.3$&$2.316 \pm 0.004$ & $54.9 \pm 1.3$&$
38.3 \pm 1.0$& 107\\
4\dotfill & & 182E, 31S & $737.5  \pm 0.7$&$1.438 \pm 0.001$ & $38.5 \pm 0.2$&$
34.9 \pm 0.2$& 122\\
5\dotfill & &  98W, 86N & $731.0  \pm 1.0$&$2.269 \pm 0.003$ & $53.7 \pm 1.3$&$
39.8 \pm 1.0$& 26\\
6\dotfill & &  50W,  6S & $718.7  \pm 1.3$&$1.489 \pm 0.003$ & $45.2 \pm 0.7$&$
31.6 \pm 0.2$& 171\\
7\dotfill & &   6W, 109S& $700.7  \pm 2.0$&$2.834 \pm 0.001$ & $78.3 \pm 2.0$&$
35.6 \pm 0.9$& 26\\
\enddata
\tablenotetext{a}{The reference position (0,0) is $\alpha\rm{(J2000)}=17^{\rm
h} 23^{\rm m} 21\rlap{.}^{\rm s} 9554$, $\delta\rm{(J2000)}=-00^{\circ} 17'
00\rlap{.}'' 938$.}
\tablenotetext{b}{Higher than $9{\sigma}=630~{\mu}$Jy.}
\tablenotetext{c}{At half maximum.}
\end{deluxetable}

\begin{deluxetable}{ccccccccc}
\tablecolumns{9}
\tablewidth{0pc}
\tablecaption{P{\footnotesize ARAMETERS OF THE} H {\footnotesize I}
C{\footnotesize LOUDS}}
\tablehead{
\colhead{Velocity\tablenotemark{a}} &
\colhead{$\Delta V_{\rm FWHM}$} &
\colhead{} &
\colhead{$\tau_{\rm peak}$} &
\colhead{} &
\colhead{$N_{\rm HI}/T_{\rm s}$} &
\colhead{$N_{\rm HI}$\tablenotemark{b}} &
\colhead{$A_{\rm v}/T_{\rm s}$\tablenotemark{c}} &
\colhead{$A_{\rm v}$\tablenotemark{b}}
\\
\colhead{(km s$^{-1}$)} &
\colhead{(km s$^{-1}$)}  &
\colhead{} &
\colhead{} &
\colhead{} &
\colhead{(cm$^{-2}~{\rm K}^{-1}$)} &
\colhead{(cm$^{-2}$)} &
\colhead{(mag~K$^{-1}$)} &
\colhead{(mag)}
\\
\colhead{(1)} &
\colhead{(2)} &
\colhead{} &
\colhead{(3)} &
\colhead{} &
\colhead{(4)} &
\colhead{(5)} &
\colhead{(6)} &
\colhead{(7)}}

\startdata
12653.5 \dotfill & 174&  &0.3&  &10.1$\times 10^{19}$ & 10.1$\times 10^{21}$
&0.066 & 6.6\\
12825.7 \dotfill & 58&  &2.5&  &28.0$\times 10^{19}$ & 28.0$\times 10^{21}$
&0.174 & 17.4\\
12883.1 \dotfill & 232&  &0.3&  &13.5$\times 10^{19}$ &13.5$\times 10^{21}$
&0.085 & 8.5\\
12940.4 \dotfill & 58&  &1.3&  &14.4$\times 10^{19}$ &14.6$\times 10^{21}$
&0.094 & 9.4\\
12997.8 \dotfill & 174&  &0.5&  &16.8$\times 10^{19}$ &16.8$\times 10^{21}$
&0.111 & 11.1 \\
\\ \enddata \tablenotetext{a}{Heliocentric velocity of the H {\footnotesize I}
absorption feature at $\tau_{\rm peak}$.}
\tablenotetext{b}{Based on $T_{\rm s}=100~\rm {K}$.}
\tablenotetext{c}{Assumes a Galactic dust-to-gas ratio.}
\end{deluxetable}

\end{document}